# The quintessence scalar field in the relativistic theory of gravity.

Yu.V.Chugreev, M.A. Mestvirishvili, K.A. Modestov

The variant of the quintessence theory is proposed in order to get an accelerated expansion of the Friedmannian Universe in the frameworks of relativistic theory of gravitation. The substance of quintessence is built up the scalar field of dark energy. It is shown, that function $V(\Phi)$, which factorising scalar field Lagrangian ($\Phi$ is a scalar field) has no influence on the evolution of the Universe. Some relations, allowing to find explicit dependence $\Phi$ on time, were found, provided given function $V(\Phi)$.

## 1. Introduction

On the edge between 20th and 21st centuries an outstanding discovery [1] had been made: it was established that expansion of the Universe presently was going with acceleration. To explain this phenomenon, it is sufficient to assume, that in the homogeneous and isotropic Universe (Friedmann Universe) there is a matter having an unusual equation of state:

$$P_Q = -(1-\nu)\rho_Q, \quad (1.1)$$

where $P_Q$ is the isotropic pressure, $\rho_Q$ is the mass density, and the following restriction [3,4] for parameter $\nu$ is a consequence of general statements of relativistic theory of gravitation (RTG) [2]:

$$0 < \nu < \frac{2}{3}. \quad (1.2)$$

A substance having such an equation of state has been called the "quintessence". It is supposed, that the quintessence is a real scalar field. No wonder, that modeling of the perfect fluid with negative preassure is provided with a rather strange scalar field Lagrangian.

Let us designate the scalar field of quintessence as $\Phi(x^\alpha)$ and postulate, that its Lagrangian density is given as follows

$$L = -\sqrt{-g}V(\Phi)(I^2)^q, \quad (1.3)$$

where $g = \det g_{\mu\nu}$, $g_{\mu\nu}$ is a metric tensor of the effective Riemannian space [2], $q$ is a number, $V(\Phi) > 0$ is some function of the field $\Phi$, and

$$I = g^{\alpha\beta}\partial_\alpha\Phi\partial_\beta\Phi \equiv g^{\alpha\beta}\nabla_\alpha\Phi\nabla_\beta\Phi, \quad (1.4)$$

In (1.4) $\nabla_\alpha$ is a covariant derivative with respect to metric $g_{\alpha\beta}$.

The Lagrangian densities of the form

$$L = -\sqrt{-g}V(\Phi)F(I), \quad (1.5)$$

where $F(I)$ is an arbitrary function of $I$, have been considered earlier [5].[1] Our choice $F(I) = (I^2)^q$ follows from intention to get equation of state (1.1) in the frameworks of the field theory of quintessence, which in their turn can explain the accelerated expansion of the Friedmannian Universe in RTG [3,4].

## 2. The equation of motion for the field $\Phi(x^\alpha)$.

According to the variational principle, the equation of motion for the field $\Phi$ which has the Lagrangian (1.3) can be obtained from the Euler-Lagrange formula

---

[1] Earlier references on this subject can be found in [5]. In the papers the field $\Phi(x^\alpha)$ with the Lagrangian (1.5) is called «κ-essence». Nevertheless, we shall keep the term "field theory of quintessence" for any field theory, resulting in (1.1), cause it is equation of state (1.1) which was corresponded to the term "quintessence" origin.

$$\frac{\partial L}{\partial \Phi} - \partial_\mu \frac{\partial L}{\partial(\partial_\mu \Phi)} = 0. \tag{2.1}$$

After substitution of Eq. (1.3) into Eq. (2.1) we obtain

$$(4q-1)\frac{d\ln V(\Phi)}{d\Phi}I^2 - 8q(2q-1)g^{\alpha\mu}\partial_\alpha\Phi(\Gamma^\lambda_{\mu\tau}g^{\tau\beta}\partial_\lambda\partial_\beta\Phi - g^{\lambda\beta}\partial_\beta\Phi\partial_\lambda\partial_\mu\Phi) -$$

$$-4q(g^{\mu\tau}\Gamma^\alpha_{\mu\tau}\partial_\alpha\Phi - g^{\lambda\mu}\partial_\lambda\partial_\mu\Phi)I = 0. \tag{2.2}$$

Here we have

$$\Gamma^\alpha_{\mu\tau} = \frac{1}{2}g^{\alpha\beta}(\partial_\mu g_{\beta\tau} + \partial_\tau g_{\beta\mu} - \partial_\beta g_{\mu\tau}).$$

In terms of covariant derivatives $\nabla_\alpha$, Eq. (2.2) has the following form

$$(4q-1)\frac{d\ln V(\Phi)}{d\Phi}I^2 + 8q(2q-1)g^{\alpha\mu}g^{\lambda\beta}\nabla_\alpha\Phi\nabla_\beta\Phi\nabla_\lambda\nabla_\mu\Phi +$$

$$+4qg^{\lambda\mu}\nabla_\lambda\nabla_\mu\Phi I = 0. \tag{2.3}$$

From the Lagrangian density (1.3) we can derive the symmetric energy-momentum tensor density of the field $\Phi(x^\alpha)$. According to Hilbert,

$$T_{\mu\nu} = 2\frac{\delta L}{\delta g^{\mu\nu}}.$$

As $L$ does not depend on derivatives of the metric tensor $g_{\alpha\beta}$, we obtain

$$T_{\mu\nu} = 2\frac{\partial L}{\partial g^{\mu\nu}} = \sqrt{-g}V(\Phi)(I^2)^{q-1}[g_{\mu\nu}I^2 - 4qI\partial_\mu\Phi\partial_\nu\Phi]. \tag{2.4}$$

### 3. Equation of motion for field $\Phi$ and its energy-momentum tensor in the homogeneous and isotropic Universe.

For a homogeneous and isotropic Universe metric tensors $g_{\alpha\beta}$ and $g^{\alpha\beta}$ in Cartesian coordinates are given as follows [2]:

$$g_{00} = 1;\ g_{11} = g_{22} = g_{33} = -a_{max}^4 a^2,\ g_{\alpha\beta} = 0,\ \alpha \neq \beta \tag{3.1}$$

and

$$g^{00} = 1;\ g^{11} = g^{22} = g^{33} = -\frac{1}{a_{max}^4 a^2},\ g^{\alpha\beta} = 0.\ \alpha \neq \beta \tag{3.1'}$$

Here $a(\tau)$ is the scale factor, and $a_{max} < \infty$ is its maximal value. The square root $\sqrt{-g}$ is equal to

$$\sqrt{-g} = a_{max}^6 a^3. \tag{3.2}$$

To provide that field $\Phi(x^\alpha)$ does not break homogeneous and isotropic character of the Universe, it is necessary to assume, that $\Phi(x^\alpha)$ depends on $\tau$ only. Then, according to (1.4) and (3.1'),

$$I = (\partial_0\Phi)^2$$

and for $T_{\mu\nu}$ taken from Eq. (2.4), in view of Eq. (3.1), we obtain

$$T_{00} = (1-4q)a_{max}^6 a^3 V(\Phi)[(\partial_0\Phi)^2]^{2q}, \tag{3.3}$$

$$T_{11} = T_{22} = T_{33} = -a_{max}^{10} a^5 V(\Phi)[(\partial_0\Phi)^2]^{2q}, \tag{3.4}$$

$$T_{\mu\nu} = 0, \quad \mu \neq \nu.$$

It is standardly assumed that in the Friedman Universe the matter can be described by the energy-momentum tensor density of some perfect fluid. This tensor is written as follows

$$T_{\mu\nu} = \sqrt{-g}[(\rho + p)U_\mu U_\nu - g_{\mu\nu} p]. \qquad (3.5)$$

As for the homogeneous and isotropic Universe 4-velocity $U_\mu$ has the form:

$$U_0 = 1, \quad U_k = 0,$$

we find from Eq. (3.5), in view of Eq. (3.1),

$$T_{00} = a_{max}^6 a^3 \rho, \qquad (3.6)$$

$$T_{11} = T_{22} = T_{33} = a_{max}^{10} a^5 p, \quad T_{\mu\nu} = 0, \quad \mu \neq \nu. \qquad (3.7)$$

If we extend this hypothesis of an opportunity to represent the energy-momentum tensor density of matter by the ideal fluid energy-momentum tensor density, onto the quintessence field, then after comparison Eq. (3.3) and Eq. (3.4) with Eq. (3.6) and Eq. (3.7) accordingly, we obtain

$$\rho_Q = (1 - 4q)V(\Phi)[(\partial_0 \Phi)^2]^{2q}, \qquad (3.8)$$

$$P_Q = -V(\Phi)[(\partial_0 \Phi)^2]^{2q}. \qquad (3.9)$$

Hence we have

$$P_Q = -\frac{1}{(1-4q)} \rho_Q. \qquad (3.10)$$

By putting

$$1 - 4q = \frac{1}{1-\nu}, \qquad (3.11)$$

we see, that Eq. (3.10) transforms into Eq. (1.1). It gives a ground to treat field $\Phi$ with Lagrangian (1.3) as quintessence field, if only

$$q = -\frac{\nu}{4(1-\nu)}. \qquad (3.12)$$

Considering restriction (1.2) for parameter $\nu$, we obtain from Eq. (3.12)

$$-\frac{1}{2} < q < 0. \qquad (3.13)$$

Thus, the Lagrangian density for quintessence has the following form

$$L = -\sqrt{-g} V(\Phi)(I^2)^{-\frac{\nu}{4(1-\nu)}}, \qquad (3.14)$$

and the density and pressure for this field in the homogeneous and isotropic Universe will be given as follows

$$\rho_Q = \frac{1}{1-\nu} V(\Phi)[(\partial_0 \Phi)^2]^{-\frac{\nu}{2(1-\nu)}}, \qquad (3.15)$$

$$P_Q = -V(\Phi)[(\partial_0 \Phi)^2]^{-\frac{\nu}{2(1-\nu)}}. \qquad (3.16)$$

To obtain explicit form of (3.15) and (3.16), it is necessary to solve equation of motion (2.2). As for the Friedmann Universe we have Christoffel symbols of the form:

$$\Gamma^0_{00} = 0, \quad \Gamma^0_{ik} = a_{max}^4 a \partial_0 a \delta_{ik}; \quad g^{\mu\tau} \Gamma^0_{\mu\tau} = -\frac{3}{a} \partial_0 a,$$

then from Eq. (2.2), in view of Eq. (3.12) we find

$$-\frac{1}{1-\nu}\frac{d\ln V(\Phi)}{d\Phi}(\partial_0\Phi)^4 - \frac{3\nu}{(1-\nu)}\frac{1}{a}\partial_0 a(\partial_0\Phi)^3 +$$

$$+\frac{\nu}{(1-\nu)^2}(\partial_0\Phi)^2\partial_0^2\Phi = 0.$$

By dividing this expression over $-\frac{1}{(1-\nu)^2}(\partial_0\Phi)^3$ and considering identity

$$\frac{d\ln V(\Phi)}{d\Phi}\partial_0\Phi \equiv \partial_0\ln V(\Phi),$$

we obtain

$$\partial_0\ln\left[(V(\Phi))^{1-\nu}|\partial_0\Phi|^{-\nu} a^{3\nu(1-\nu)}\right] = 0,$$

and then

$$V(\Phi)|\partial_0\Phi|^{-\frac{\nu}{1-\nu}} = \frac{\Lambda}{a^{3\nu}}, \quad \Lambda = const > 0. \tag{3.17}$$

By substituting this expression into Eq. (3.15) and Eq. (3.16), we finally get

$$\rho_Q = \frac{\Lambda}{1-\nu}\frac{1}{a^{3\nu}}, \tag{3.18}$$

$$P_Q = -\frac{\Lambda}{a^{3\nu}}. \tag{3.19}$$

It is well-know [2,4], that for equation of state (1.1) the covariant conservation law $\nabla_\mu T^{\mu\nu} = 0$, where $T^{\mu\nu}$, looks like (3.5), in the Friedmann Universe leads to formulas

$$\rho_Q = \frac{B}{a^{3\nu}} \quad and \quad P_Q = -(1-\nu)\frac{B}{a^{3\nu}}, \quad where \quad B = const > 0. \tag{3.20}$$

After a redefinition of constant $\Lambda$ as follows

$$\Lambda = (1-\nu)B$$

we can see, that Eqs. (3.18) - (3.19) transform into (3.20), that means the equivalence of the equation of motion (2.3) to the covariant conservation law $\nabla_\mu T^{\mu\nu}$, when the last one is supplemented by Eq. (1.1).

However, it is important to note, that for deriving basic formulas (3.18) and (3.19), describing the evolution of homogeneous and isotropic Universe, the explicit form of $V(\Phi)$ is not required. Let's denote the values of the scaling factor $a(\tau)$ and the scalar field $\Phi(\tau)$ at the start time of expansion of the Universe $\tau = 0$ as $a_{min}$ and $\Phi(0)$ correspondingly. Then Eq. (3.17) can be rewritten in the form

$$\int_{\Phi(0)}^{\Phi(\tau)}\frac{d\Phi}{V(\Phi)^{\frac{1-\nu}{\nu}}} = \Lambda^{-\frac{(1-\nu)}{\nu}}\int_0^\tau a^{3(1-\nu)}d\tau, \tag{3.21}$$

The Friedmannian Universe in the RTG performs the cyclic motion, with the scaling factor $a(\tau)$ changing in the domain

$$0 < a_{min} \le a(\tau) \le a_{max} < \infty. \tag{3.22}$$

Therefore the integral

$$J(\tau) = \int_0^\tau a^{3(1-\nu)} d\tau \qquad (3.23)$$

is a finite one. Consequently, the l.h.s. of the Eq.(3.21) should be finite as well. It imposes a (very weak) limit on the function $V(\Phi)$.

With the help of RTG results one can express the integral (3.23) through the observables in order to lighten the further analisis. Toward this end it is possible to rewright the Eq.(3.23) as follows

$$J(\tau) = \int_{a_{min}}^{a(\tau)} a^{2-3\nu}\left(a \frac{d\tau}{da}\right) da . \qquad (3.24)$$

According RTG [2,4] for the expansion stage of the Universe evolution we have

$$a \frac{d\tau}{da} = \left[\frac{8\pi G \rho}{3} - \frac{m^2}{6}\left(1 - \frac{3}{2\beta^4 a^2} + \frac{1}{2a^6}\right)\right]^{-\frac{1}{2}}, \qquad (3.25)$$

where $\beta = a_{max}$, $m^2 = \left(\frac{m_g c^2}{\hbar}\right)^2$, $m_g$ is the rest mass of the graviton ($m_g \cong 3.6 \cdot 10^{-66} g$ [4]), and $\rho$ is a total mass density in the Universe:

$$\rho = \rho_r + \rho_m + \rho_Q . \qquad (3.26)$$

Here $\rho_r$ is a mass density at the radiation stage, $\rho_m$ is a mass density at the barion domination's phase, and $\rho_Q$ is a mass density at the quintessence period.

Let's introduce the following notations:

I. $x = \frac{a(\tau)}{a_0}$, where $a_0$ is a value of the scaling factor $a(\tau)$ at the present moment $\tau = \tau_{pres}$. The order of magnitude of $a_0$ is [4] $a_0 \approx 5 \cdot 10^9$.

II. The Hubble constant

$$H = \left(\frac{1}{a}\frac{da}{d\tau}\right)_{\tau = \tau_{pres}}$$

and the critical density $\rho_c^0$:

$$H^2 = \frac{8\pi G}{3} \rho_c^0$$

III. Relative densities

$$\Omega_r^0 = \frac{\rho_r^0}{\rho_c^0}; \quad \Omega_m^0 = \frac{\rho_m^0}{\rho_c^0} \quad \text{and} \quad \Omega_Q^0 = \frac{\rho_Q^0}{\rho_c^0},$$

where $\rho_r^0$, $\rho_m^0$ and $\rho_Q^0$ are the densities of the correspondingly radiation-, barion- and quintessence stages at the present moment $\tau = \tau_{pres}$.

Since [2,4]

$$\rho_r = \frac{A_r}{a^4}; \quad \rho_m = \frac{A_m}{a^3} \quad \text{и} \quad \rho_Q = \frac{A_Q}{a^{3\nu}},$$

where $A_r$, $A_m$ и $A_Q$ are some constants, then in terms of the variable $x$ we have:

$$\rho_r = \frac{\rho_r^0}{x^4}, \quad \rho_m = \frac{\rho_m^0}{x^3} \quad \text{и} \quad \rho_Q = \frac{\rho_Q^0}{x^{3\nu}}.$$

Therefore

$$\frac{8\pi G}{3}\rho = H^2\left(\frac{\Omega_r^0}{x^4} + \frac{\Omega_m^0}{x^3} + \frac{\Omega_Q^0}{x^{3\nu}}\right) \quad (3.27)$$

Taking into account Eq. (3.25) and Eq.(3.27) one can get the integral $J(\tau)$ in the form:

$$J(\tau) = \frac{a_0^{3(1-\nu)}}{H}\int_{x_{min}}^{x(\tau)} x^{2-3\nu}\left[\frac{\Omega_r^0}{x^4} + \frac{\Omega_m^0}{x^3} + \frac{\Omega_Q^0}{x^{3\nu}} - \frac{f^2}{6}\left(1 - \frac{3}{2\beta^4 a_0^2 x^2} + \frac{1}{2a_0^6 x^6}\right)\right]^{-\frac{1}{2}} dx, \quad (3.28)$$

where

$$f = \frac{m}{H} = \frac{m_g c^2}{\hbar H} \quad \text{and} \quad x_{min} = \frac{a_{min}}{a_0}. \quad (3.29)$$

The r.h.s. of the Eq.(3.28) contains observable parameters [4]. In particular, present values of relative densities are [6]

$$\Omega_r^0 \approx 0.6\cdot 10^{-4}; \quad \Omega_m^0 \approx 0.3 \text{ и } \Omega_Q^0 \approx 0.7. \quad (3.30)$$

At the domination-of- radiation stage

$$a_{min} \leq a(\tau) \leq a_r,$$

(where $a_r$ is the value of the scaling factor when in the square root of Eq.(3.28) the terms $\Omega_r^o$ and $f^2/12a_0^6 x^6$ prevail over the rest ones), one can get the following approximation for $J(\tau)$:

$$J_r(\tau) \cong \frac{a_0^{3(1-\nu)}}{H\sqrt{\Omega_r^0}}\int_{x_{min}}^{x(\tau)} \frac{x^{5-3\nu} dx}{\sqrt{x^2 - x_{min}^2}}, \quad (3.31)$$

where

$$x_{min} \equiv \frac{a_{min}}{a_0} = \left(\frac{f^2}{12 a_0^6 \Omega_r^0}\right)^{\frac{1}{2}}, \quad (3.32)$$

and

$$x(\tau) \leq x_r = \left(\frac{a_r}{a_0}\right).$$

After integrating Eq.(3.31) we obtain

$$J_r(\tau) \cong \frac{a_{min}^{3(1-\nu)}}{H\sqrt{\Omega_r^0}} x_{min}^2 \left[\left(\frac{a}{a_{min}}\right)^2 - 1\right]^{\frac{1}{2}} F\left(-2 + \frac{3\nu}{2}, \frac{1}{2}; \frac{3}{2}; 1 - \left(\frac{a}{a_{min}}\right)^2\right). \quad (3.33)$$

In small vicinity of $a_{min}$ $J_r(\tau)$ is more simple

$$J_r(\tau) \cong \frac{a_{min}^{3(1-\nu)}}{H\sqrt{\Omega_r^0}} x_{min}^2 \left[\left(\frac{a}{a_{min}}\right)^2 - 1\right]^{\frac{1}{2}}. \quad (3.34)$$

Let's calculate $J(\tau)$ for the quintessence period. New relative variable is given as follows $x_Q = \frac{a_Q}{a_0}$, where $a_Q$ is the value of scaling factor at the moment, when the quintessence starts to dominate in the Universe.

Taking into account strong ineqalities

$$\frac{\Omega_r^0}{x^4} \ll \frac{\Omega_Q^0}{x^{3\nu}} \quad \text{и} \quad \frac{\Omega_m^0}{x^3} \ll \frac{\Omega_Q^0}{x^{3\nu}}$$

and Eq. (3.30) one can easily get the following estimation

$$\left(\frac{3}{7}\right)^{\frac{1}{3(1-\nu)}} < 1 << x_Q \le x_{max}. \tag{3.35}$$

We can now rewrite (3.28) in the form

$$J(\tau) \approx J_1 + \frac{a_0^{3(1-\nu)}\sqrt{6}}{Hf}\int_{x_Q}^{x(\tau)}\frac{x^{2-\frac{3\nu}{2}}dx}{\sqrt{\frac{\Omega_Q^0 6}{f^2}-x^{3\nu}}}, \tag{3.36}$$

where

$$J_1 = \frac{a_0^{3(1-\nu)}}{H}\int_{x_{min}}^{x_Q}\frac{x^{2-3\nu}dx}{\left[\frac{\Omega_r^0}{x^4}+\frac{\Omega_m^0}{x^3}+\frac{\Omega_Q^0}{x^{3\nu}}-\frac{f^2}{6}\left(1-\frac{3}{2\beta^4 a_0^2 x^2}+\frac{1}{2a_0^6 x^6}\right)\right]^{\frac{1}{2}}}. \tag{3.37}$$

Integral in (3.36)

$$I = \int_{x_Q}^{x(\tau)}\frac{x^{2-\frac{3\nu}{2}}dx}{\sqrt{\frac{\Omega_Q^0 6}{f^2}-x^{3\nu}}}$$

equals to

$$I \approx \frac{2}{3(2-\nu)}\left(\frac{a(\tau)}{a_0}\right)^{3\left(1-\frac{\nu}{2}\right)}\frac{1}{\sqrt{t_{max}}}\left[F\left(\frac{1}{2},\frac{1}{\nu}-\frac{1}{2};\frac{1}{\nu}+\frac{1}{2};\frac{a^{3\nu}}{a_0^{3\nu}t_{max}}\right)-\left(\frac{a_Q}{a(\tau)}\right)^{3\left(1-\frac{\nu}{2}\right)}\right], \tag{3.38}$$

where

$$t_{max} \equiv \left(\frac{a_{max}}{a_0}\right)^{3\nu} = \frac{\Omega_Q^0 6}{f^2}. \tag{3.39}$$

By substituting (3.38) in (3.36), we obtain the value of $J(\tau)$ for the quintessence stage:

$$J(\tau) \approx J_1 + \frac{2}{3(2-\nu)H\sqrt{\Omega_Q^0}}\left(\frac{a(\tau)}{a_0}\right)^{3\left(1-\frac{\nu}{2}\right)}\left[F\left(\frac{1}{2},\frac{1}{\nu}-\frac{1}{2};\frac{1}{\nu}+\frac{1}{2};\frac{a^{3\nu}}{a_0^{3\nu}t_{max}}\right)-\left(\frac{a_Q}{a(\tau)}\right)^{3\left(1-\frac{\nu}{2}\right)}\right]. \tag{3.40}$$

In order to get the dependance $J = J(\tau)$, it is necessary to find the c, i.e. to solve Eq. (3.25). With following [4], let's express it in terms of $I - III$. Then solution will have the form

$$\tau = \frac{1}{H}\int_{x_{min}}^{x(\tau)}\frac{dx}{x\left[\frac{\Omega_r^0}{x^4}+\frac{\Omega_m^0}{x^3}+\frac{\Omega_Q^0}{x^{3\nu}}-\frac{f^2}{6}\left(1-\frac{3}{2\beta^4 a_0^2 x^2}+\frac{1}{2a_0^6 x^6}\right)\right]^{\frac{1}{2}}}. \tag{3.41}$$

The Eq. (3.41) was analysed in [4], provided in the vicinity of $\tau \approx 0$

$$a(\tau) \approx a_{min}\left(1+\frac{4\pi G}{3}\rho_{max}\tau^2\right). \tag{3.42}$$

In those part of Universe evolution, where one can neglect massive terms (proportional to $f$ in the integral (3.41)), the function $a = a(\tau)$ is well known. It coincides with that one in General Relativity.

When the quintessence dominates in the Universe, and $a(\tau) \gg 1$, the Eq. (3.41) is more simple:

$$\tau \approx \tau_1 + \frac{\sqrt{6}}{Hf} \int_{x_Q}^{x(\tau)} \frac{x^{\frac{3\nu}{2}-1} dx}{\sqrt{\frac{\Omega_Q^0 6}{f^2} - x^{3\nu}}}. \tag{3.43}$$

Here

$$\tau_1 = \frac{1}{H} \int_{x_{\min}}^{x_Q} \frac{dx}{x \left[ \frac{\Omega_r^0}{x^4} + \frac{\Omega_m^0}{x^3} + \frac{\Omega_Q^0}{x^{3\nu}} - \frac{f^2}{6}\left(1 - \frac{3}{2\beta^4 a_0^2 x^2} + \frac{1}{2 a_0^6 x^6}\right) \right]^{\frac{1}{2}}} \tag{3.44}$$

-moment of the quintessence domination in the Universe, with $\tau_1$ being less than the present time $\tau_1 \leq \tau_{pres}$.

Since

$$\int_{x_Q}^{x(\tau)} \frac{x^{\frac{3\nu}{2}-1} dx}{\sqrt{\frac{\Omega_Q^0 6}{f^2} - x^{3\nu}}} \approx \frac{1}{3\nu} \arccos\left(1 - 2\frac{x^{3\nu}}{t_{\max}}\right) - \frac{2}{3\nu} \frac{x_Q^{3\nu}}{\sqrt{t_{\max}}},$$

then puting this relation into the Eq. (3.43), one can find

$$\left(\frac{a(\tau)}{a_0}\right)^{3\nu} \approx \frac{3\Omega_Q^0}{f^2} \left[ 1 - \cos\left(\sqrt{\frac{3}{2}} f \left( H\nu(\tau - \tau_1) + \frac{2}{3}\left(\frac{a_Q}{a_0}\right)^{3\nu} \frac{1}{\sqrt{\Omega_Q^0}}\right)\right)\right]. \tag{3.45}$$

As $x_{\max}^{3\nu} \equiv \left(\frac{a_{\max}}{a_0}\right)^{3\nu} = \frac{6\Omega_Q^0}{f^2}$, and for $a = a_{\max}$ we have $\tau = \tau_{\max} \gg \tau_1$, then from (3.45) it follows that

$$\sqrt{\frac{3}{2}} f H\nu \tau_{\max} \approx \pi, \tag{3.46}$$

From here one can find $\tau_{\max}$. In the vicinity of $\tau \approx \tau_{\max}$ with the help of Eq. (3.45) one can easily establish, that

$$a(\tau) \approx a_{\max} \left[1 - \frac{\pi^2}{4}\left(1 - \frac{\tau}{\tau_{\max}}\right)^2\right]^{\frac{1}{3\nu}}. \tag{3.47}$$

Putting Eq. (3.42) into Eq. (3.34), we can calculate the value of $J(\tau)$ in the vicinity of $\tau \approx 0$:

$$J_r(\tau) \approx \frac{a_{\min}^{3(1-\nu)}}{H\sqrt{\Omega_r^0}} x_{\min}^2 \left(\frac{8\pi G \rho_{\max}}{3}\right)^{\frac{1}{2}} \tau. \tag{3.48}$$

According to the mean value theorem, expression of $J_1$ (см.(3.37)) has the form

$$J_1 = (x_0 x_1)^{3(1-\nu)} \tau_1, \tag{3.49}$$

where $x_1$ belongs the interval $x_1 \in (x_{\min}, x_Q)$, with $\tau_1$ being determined by the Eq.(3.44).

Taking into account Eq. (3.39) for $t_{\max}$, and Eq. (3.47) one can find

$$\left(\frac{a}{a_0}\right)^{3\nu} \frac{1}{t_{max}} = \left(\frac{a}{a_{max}}\right)^{3\nu} \approx 1 - \frac{\pi^2}{4}\left(1 - \frac{\tau}{\tau_{max}}\right)^2. \tag{3.50}$$

Puting Eqs.(3.47), (3.49) and (3.50) into Eq. (3.40) for $J(\tau)$ in the vicinity of $\tau \approx \tau_{max}$, we obtain finally

$$J(\tau) \approx (x_0 x_1)^{3(1-\nu)} \tau_1 +$$

$$+ \frac{2}{3(2-\nu)H\sqrt{\Omega_Q^0}} \left(\frac{a_{max}}{a_0}\right)^{3\left(1-\frac{\nu}{2}\right)} \left[1 - \frac{\pi^2}{4}\left(1 - \frac{\tau}{\tau_{max}}\right)^2\right]^{\frac{2-\nu}{2\nu}} F\left(\frac{1}{2}, \frac{1}{\nu} - \frac{1}{2}; \frac{1}{\nu} + \frac{1}{2}; 1 - \frac{\pi^2}{4}\left(1 - \frac{\tau}{\tau_{max}}\right)^2\right). \tag{3.51}$$

One can get the explicit form of the field $\Phi(\tau)$, provided the function $V(\Phi)$ is known. The Friedmannian model itself does not impose any constraints on the field. In other words, different quintessence theories with different Lagrangians give the same values of $\rho_Q$ and $p_Q$. Consequently, all of them determine the same time evolution of the Universe.

In some papers their authors have chose power low function $V(\Phi) \sim \Phi^{-\alpha}$, where $\alpha$ -some positive number. Evidently, that in this case one can express $\Phi(\tau)$ through $J(\tau)$ explicitly using Eq. (3.21).

### 4. Acknowledgments

At this end the authors express their deep gratitude to A.A.Logunov for permanent interes to this work and numerous discussions. One of us (M.A.M.) would like to acknowledge colleagues from Department of Theoretical Physics of the Institute of High Energy Physics  V.V.Kiselev, A.V.Razumov, A.P.Samokhin, V.O.Soloviev, Yu.G.Stroganov, Yu.M.Zinoviev, for helpful discussions.

### References

1. Ries A.G. et al.  Astron.J. 1988. v.116. p.1009-1038;
   Perlmutter S. et al.  Nature. 1988. v.391. p.51-54;
   Astrophys. J. 1999. v.517. p.565-586;
   Bennett C.L. et al.  Astrophys. J. 1996. v.64. p.L1-L4;
   Hanany S. et al.  Astrophys. J. 2000. v.545. p.L5-L9;
   Bernardis P. et al.  Nature. 2000. v.404.p.955-959;
   Benoit A. et al.  Astronomy and Astrophysics. 2003.v.399.L19-L23; L25-L30;
   Bennet C.L. et al.  Astrophys. J.S. 2003. v.148. p.1-28;
   Spergel D.N. et al.  Astrophys. J.S. 2003. v.148. p.175-194;
   Percival W.J. et al.   MNRAS. 2001.v.327. p.1297-1306;
   Verde L. et al.   MNRAS. 2002.v.335. p.432-440;
   Verde L. et al.   Astrophys. J.S. 2003. v.148. p.195-212;
   York D.G. et al.   Astron. J. 2000. v.120. p.1579-1587;
   Stroughton C. et al.   Astron. J. 2002. v.123. p.485-548;
   Abazajian K. et al.   Astron. J. 2000. v.126. p.2081-2086.
2. Logunov A.A. The Thory of Gravity.- Moscow:Nauka, 2001; gr-qc/0210005.
3. Kalashnikov V.L. gr-qc/0109060; gr-qc/0202084.
4. Gershtein S.S., Logunov A.A., Mestvirishvili M.A. and Tkachenko N.P. Fiz. Elem. Chast. Atomn. Yadra 2005.V.36. Issue5. P.1003-1050.
5. Chimento L.P. et al.   Mod.Phys.Lett.A. 2004. v.19. p.761-768;
   Scherrer R.J. et al. Phys.Rev.Lett. 2004. v.93. p.011301-4;
   Rupam D. et al.   gr-qc/0609014.
6. Tegmark M. et al.    Phys.Rev.D. 2004 v.69.p.103501-26.